\documentclass[twocolumn,prl,showpacs,superscriptaddress,]{revtex4}
\usepackage{amssymb}
\usepackage{amsmath}
\usepackage{graphicx}
\usepackage{amsfonts}

\begin{document}

\title{Maxwell's Demon Assisted Thermodynamic Cycle in Superconducting
Quantum Circuits}
\author{H.T. Quan}
\affiliation{Frontier Research System, The Institute of Physical and Chemical Research
(RIKEN), Wako-shi, Saitama 351-0198, Japan }
\affiliation{Institute of Theoretical Physics, Chinese Academy of Sciences, Beijing,
100080, China}
\author{Y.D. Wang}
\affiliation{Frontier Research System, The Institute of Physical and Chemical Research
(RIKEN), Wako-shi, Saitama 351-0198, Japan }
\affiliation{Institute of Theoretical Physics, Chinese Academy of Sciences, Beijing,
100080, China}
\author{Yu-xi Liu}
\affiliation{Frontier Research System, The Institute of Physical and Chemical Research
(RIKEN), Wako-shi, Saitama 351-0198, Japan }
\author{C.P. Sun}
\affiliation{Frontier Research System, The Institute of Physical and Chemical Research
(RIKEN), Wako-shi, Saitama 351-0198, Japan }
\affiliation{Institute of Theoretical Physics, Chinese Academy of Sciences, Beijing,
100080, China}
\author{Franco Nori}
\affiliation{Frontier Research System, The Institute of Physical and Chemical Research
(RIKEN), Wako-shi, Saitama 351-0198, Japan }
\affiliation{Center for Theoretical Physics, Physics Department, CSCS, The University of
Michigan, Ann Arbor, Michigan 48109-1040}
\date{\today}

\begin{abstract}
We study a new quantum heat engine (QHE), which is assisted by a Maxwell's
demon. The QHE requires three steps: thermalization, quantum measurement,
and quantum feedback controlled by the Maxwell demon. We derive the
positive-work condition and operation efficiency of this composite QHE.
Using controllable superconducting quantum circuits as an example, we show
how to construct our QHE. The essential role of the demon is explicitly
demonstrated in this macroscopic QHE.
\end{abstract}

\pacs{05.30.-d, 03.67.-a, 85.25.Cp}
\maketitle

\emph{Introduction.---} A Maxwell demon is a construct that can
distinguish the velocities of individual gas molecules and then
separate hot and cold molecules into two domains of a container,
so that the two domains will have different
temperatures~\cite{demon}. This result seems to contradict the
second law of thermodynamics, because one can put a heat engine
between them to extract work. The solution of this
puzzle~\cite{demon} refers to the so-called Landauer's
principle~\cite{Landauer,bennett} that essentially links
information theory with fundamental physics~\cite{maruyama}.
Several quantum heat engines (QHEs) assisted by Maxwell's demons
have been proposed in Refs.~\cite{lloyd,scully,zurek}.

Here, we propose a new QHE model integrated with a built-in quantum
Maxwell's demon performing both: the quantum measurement on the working
substance, and the feedback control for the system according to the
measurement. We demonstrate the role of Maxwell's demon in a fully quantum
manner. The thermodynamic cycle in our setup contains three fundamental
stages: (i) a CNOT operation, making a pre-measurement to extract
information from the working substance; (ii) the feedback-action of the
demon controlling the working substance to extract work; and (iii) the
disentanglement process that thermalizes the working substance and the demon
by two separate thermal baths. The demon plays a role in the first two steps.

We further illustrate how to implement our QHE using superconducting qubit
circuits \cite{wang,liu}. In our setup, the demon-assisted working substance
does work via two CNOT operations, which can be realized by single-qubit
operations and an easily realized i-SWAP operation. The CNOT operation
performs the basic functions of the quantum demon.
\begin{figure}[tbp]
\includegraphics[bb=110 380 499 690, width=6.0 cm, clip]{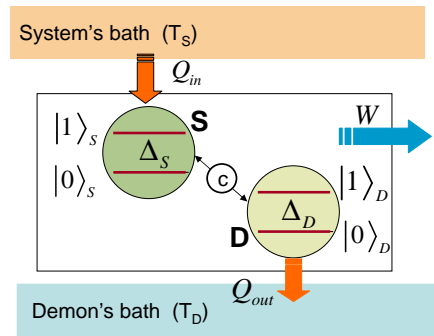}
\caption{(color online). Schematics of the Maxwell's-demon-based
quantum heat engine (QHE). The qubit $S$ is the ``working
substance" system, which is monitored and then controlled by
another qubit $D$, acting as a Maxwell's demon. The central
circle with the letter ``c" denotes a switchable coupling
between the $S$ and the $D$. $S$ \textit{plus} $D$ form the QHE.
$Q_{\rm in}$ and $Q_{\rm out}$ indicate the heat absorbed and
released; $W$ denotes the work done. When the erasure of $D$ is
included in the cycle, according to Landauer's principle, the
violation of the Second Law is prevented.} \label{fig.1}
\end{figure}

\emph{Maxwell's demon-assisted thermodynamic cycle in two-qubit
system.---} Our QHE cycle is similar to a quantum Otto
cycle~\cite{explanation} described in Ref.~\cite{kieu} and
generalized in Ref.~\cite{quan}. Here, the QHE, shown in
Fig.~\ref{fig.1}, is a composite system consisting of two
qubits: the ``working substance" $S$ and the quantum Maxwell's
demon $D$. They are separately coupled to two different heat
baths with the temperatures $T_{S}$ and $T_{D}$. Using the Pauli
matrices $\sigma _{\alpha }^{(F)}$ $(F=S$, $D;\alpha =x,y,z)$,
the model Hamiltonian can be written as
\begin{equation}
H_{I}=\sum_{F=S,D}\Delta _{F}\,\sigma
_{z}^{(F)}+E_{L}\mathbf{\,}(\sigma _{x}^{(S)}\sigma
_{x}^{(D)}-\sigma _{y}^{(S)}\sigma _{y}^{(D)}),  \label{0}
\end{equation}
where $\Delta_{F}$ is the level spacing of the qubits and
$E_{L}$ is a controllable coupling strength between $S$ and $D$.
Using both the controllable $XY$ interaction and on-site
potentials in Eq.~(\ref{0}),  we can realize various quantum
logic operations~\cite{schuch}. Now, let us study each step of
our QHE cycle and calculate the work done and the heat absorbed
in each step of the thermodynamic cycle.

\emph{S1}: $S$ and $D$ are decoupled by setting $E_{L}=0$ in
Eq.~(\ref{0}) and separately coupled to two heat baths with
different temperatures $T_{S}$ and $T_{D}$. As shown below,
whatever are the initial states of $S$ and $D$, either entangled
or separated, after a thermalization, they will reach their
respective equilibrium states: $\rho
_{F}(1)=p_{F}(0)|0_{F}\rangle \langle
0_{F}|+p_{F}(1)|1_{F}\rangle \langle 1_{F}|$, with $F=S$ or $D$.
Here, $p_{F}(1)=\exp (-\beta _{F}\Delta _{F})/z_{F}$, and
$p_{F}(0)=1/z_{F}$, are the Boltzmann probability distributions
for two energy levels; $z_{F}=1+\exp (-\beta _{F}\Delta _{F})$
is the partition function with $\beta _{F}=1/(k_{B}T_{F})$,
where $k_{B}$ is the Boltzmann constant. We have chosen the
ground state energy as zero. The thermalized state $\rho
(1)=\rho _{S}(1)\otimes \rho _{D}(1)$ of the total system is
\begin{eqnarray}
\rho (1) &=&p_{S,D}^{1,1}\,|1,1\rangle \langle
1,1|+p_{S,D}^{1,0}\,|1,0\rangle \langle 1,0|  \label{1} \\
&+&p_{S,D}^{0,1}\,|0,1\rangle \langle 0,1|+p_{S,D}^{0,0}\,|0,0\rangle
\langle 0,0|,  \notag
\end{eqnarray}%
where for $F,F^{\prime }=S,D$ and $\,q,\,q^{\prime }=0,1$, $|q,q^{\prime
}\rangle \equiv $ $|q_{S}\rangle \otimes |q_{D}^{\prime }\rangle $ and $%
p_{F,F^{\prime }}^{q,q^{\prime }}\equiv p_{F}\left( q\right) \,p_{F^{\prime
}}\left( q^{\prime }\right) $ being the joint probabilities.

\emph{S2}: The second step is a CNOT operation flipping the
demon states only when the working substance system is in its
excited state~\cite{lloyd}. In this step, the demon acquires
information about the system. This CNOT process can be
realized~\cite{schuch} by the controllable Hamiltonian
~(\ref{0}) and is assumed to be so short that the coupling of
$S$ and $D$ to the baths can be ignored. Thus, $\rho (1)$, after
the second step, is changed to
\begin{eqnarray}
\rho (2) &=&p_{S,D}^{1,1}\,|1,0\rangle \langle
1,0|+p_{S,D}^{1,0}\,|1,1\rangle \langle 1,1|  \label{2} \\
&&+p_{S,D}^{0,1}\,|0,1\rangle \langle 0,1|+p_{S,D}^{0,0}\,|0,0\rangle
\langle 0,0|.  \notag
\end{eqnarray}%
The entropy of $\rho (2)$ is equal to that of $\rho (1)$, i.e., measurements
do not lead to entropy increase \cite{Landauer,bennett,zurek}. This agrees
well with Landauer's principle.

\emph{S3}: In the third step, the demon controls the system to
do work according to the information acquired by the demon about
the system. Physically, the system experiences a conditional
evolution (CEV) $U_{c}$ which can be realized by the
Hamiltonian~(\ref{0}), that is,  $|q_{S}\rangle \otimes
|q_{D}^{\prime }\rangle \rightarrow (U_{c})^{q^{\prime
}}\left\vert q_{S}\right\rangle \otimes \left\vert q_{D}^{\prime
}\right\rangle $ and $U_{c}\left\vert q_{S}\right\rangle
=\left\vert \tilde{q}_{S}\right\rangle $. Here, $\left\vert \tilde{1}%
_{S}\right\rangle =\cos \theta \left\vert 1_{S}\right\rangle +\sin \theta
\exp (i\varphi )\left\vert 0_{S}\right\rangle $, and $\left\vert \tilde{0}%
_{S}\right\rangle =-\sin \theta \left\vert 1_{S}\right\rangle
+\cos \theta \exp (i\varphi )\left\vert 0_{S}\right\rangle $,
are the states of the working substance after the conditional
evolution; $\theta $ and $\varphi $ are real parameters. A CNOT
is a special CEV for $\theta =\pi/2$. After the third step, the
density matrix $\rho (2)$ evolves into
\begin{eqnarray}
\rho (3) &=&p_{S,D}^{1,1}\,\left\vert 1,0\right\rangle \left\langle
1,0\right\vert +p_{S,D}^{1,0}\,\left\vert \tilde{1},1\right\rangle
\left\langle \tilde{1},1\right\vert  \label{3} \\
&&+p_{S,D}^{0,1}\,\left\vert \tilde{0},1\right\rangle \left\langle \tilde{0}%
,1\right\vert +p_{S,D}^{0,0}\,\left\vert 0,0\right\rangle \left\langle
0,0\right\vert .  \notag
\end{eqnarray}

Finally, the system and the demon are decoupled by setting
$E_{L}=0$ in Eq.~(\ref{0}) and brought into contact with their
own baths again, and then a new cycle starts. For each cycle
described above, we are now able to calculate the work performed
by the heat engine as $W=-\left( E_{S}^{\prime \prime
}+E_{D}^{\prime \prime }-E_{S}-E_{D}\right) $ $=\Delta
_{S}(p_{S,D}^{1,0}-p_{S,D}^{1,0}\left\vert \left\langle
\tilde{1}\right. \left\vert 1\right\rangle \right\vert
^{2}-p_{S,D}^{0,1}\left\vert \left\langle \tilde{0}\right.
\left\vert 1\right\rangle \right\vert ^{2})+\Delta
_{D}(p_{S,D}^{1,1}-p_{S,D}^{1,0}),$
\ where $E_{S}^{\prime \prime }$ ($E_{S}$) and $E_{D}^{\prime \prime }$ ($%
E_{D}$) are the internal energies of the system and demon, respectively,
after the third (first) step. The heat absorbed by the system from the heat
bath is $Q_{\mathrm{in}}=E_{S}-E_{S}^{\prime \prime }$ = $\Delta
_{S}(p_{S,D}^{1,0}-p_{S,D}^{1,0}\left\vert \left\langle \tilde{1}\right.
\left\vert 1\right\rangle \right\vert ^{2}-p_{S,D}^{0,1}\left\vert
\left\langle \tilde{0}\right. \left\vert 1\right\rangle \right\vert ^{2})$.
Based on the above results, the operation efficiency $\eta $ can be given as
\begin{equation}
\eta =W/Q_{\mathrm{in}}=1-(\Delta _{D}/\Delta _{S})\xi  \label{6}
\end{equation}%
with
\begin{equation}
\xi =\csc ^{2}\theta \,\left( p_{S,D}^{1,1}/p_{S,D}^{1,0}-1\right) \left(
p_{S,D}^{0,1}/p_{S,D}^{1,0}-1\right) ^{-1}.  \label{7}
\end{equation}

Equation~(\ref{6}) shows that $\xi \geq 0$ (to guarantee the operation
efficiency $\eta <1$). The first factor of $\xi $ in Eq.~(\ref{7}) is
positive, while the second factor, which can be simplified to $\exp (-\beta
_{D}\Delta _{D})-1$, is negative. Thus, we can conclude that the third
factor of $\xi $ in Eq.~(\ref{7}) is negative. This results in $T_{S}\geq
T_{D}(\Delta _{S}/\Delta _{D})$, and it agrees well with the positive-work
condition~\cite{kieu,quan} for a simple quantum Otto cycle without Maxwell's
demon. This coincidence is nontrivial since here $T_{S}$ and $T_{D}$ are the
temperatures of the baths surrounding qubits $S$ and $D$ in the whole cycle.
This is different from the temperatures in Refs.~\cite{kieu,quan}, where the
two temperatures are defined by two different isochoric steps in
thermodynamic cycles.

\emph{Remarks on the QHE cycle and the roles of the quantum
Maxwell's demon.---} Let us further understand each step in the
above QHE operations. We first consider the thermalization
problems for the two qubits coupled to two separated baths,
which can be modelled as two collections of harmonic oscillators
with different temperatures, e.g., $T_{S}$ and $T_{D}$. The
baths have the average thermal excitation $n(T_{F},\omega
_{F})=1/[\exp (\beta _{F}\omega _{F})-1]$ in the mode with
frequency $\omega _{F}$ ($F=S,\,D$) of the baths. After
thermalization, the population difference of $F$ can be
calculated~\cite{orszag} as
\begin{equation}
\langle \sigma _{z}^{(F)}(t)\rangle =\frac{1}{2}\left( \langle
\sigma _{z}^{(F)}(0)\rangle M_{F}+1\right) e^{-2\gamma
_{F}t}-(1/M_{F})\;,  \label{9}
\end{equation}%
where $M_{F}=1+2\,n(T_{F},\Delta _{F})$ is time-independent. The
damping rate $\gamma _{F}$ of $F$ depends on the specific
physical realization. When $t\gg 1/\gamma _{F}$, $F$ will
approach its steady state $\rho _{F}(1)$ in Eq.~(\ref{1}) with
$\langle \sigma _{z}^{(F)}(t\rightarrow \infty )\rangle
_{s}=-1/M_{F}$. Then, we can obtain the equilibrium
distribution, $p_{F}(1)=(1-1/M_{F})/2$, of the two-level system,
using $p_{F}(1)\pm p_{F}(0)=\langle \sigma _{z}^{(F)}(t)\rangle
^{(1\mp 1)/2}$. It is crucial that the steady term $\langle
\sigma _{z}^{(F)}(t\rightarrow \infty )\rangle _{s}$ in
Eq.~(\ref{9}) is independent of the initial state, since the
initial information is erased by quantum dissipation, with damping rate $%
\gamma _{F}$. Hence, whatever initial state the total system is (e.g., an
entangled state), the final steady state of $S$ or $D$ would be in its own
thermal equilibrium state.

The CNOT operation in the step \emph{S2} can be referred to a
one-bit quantum pre-measurement on the quantum system $S$ by the
Maxwell's demon \cite{zurek}. As for the CEV in step \emph{S3},
we noticed that, when we choose i) the CEV to be a special case
$\theta =\pi /2$, i.e., a CNOT, and ii) the temperature $T_{D}$
to be so low that $\exp (-\beta _{D}\Delta _{D})\ll 1$, i.e.,
the demon is \textquotedblleft erased" nearly to its ground
state $\rho _{D}\left( 1\right) \approx \left\vert
0_{D}\right\rangle \langle 0_{D}| $ \cite{lubkin}, the
efficiency of our QHE Eq.~(\ref{6}) becomes $\eta =1-(\Delta
_{D}/\Delta _{S})$. This is exactly the efficiency of a simple
quantum Otto cycle without Maxwell's demon
~\cite{explanation,kieu,quan}. Otherwise the operation
efficiency~(\ref{6}) is less than the efficiency of a simple
quantum Otto cycle. This is because i) when $T_{D}$ is
vanishingly small, the demon can be restored to a zero-entropy
\textquotedblleft standard state" ~\cite{bennett,lubkin} to
acquire information about the system in the most efficient way,
and ii) among all CEVs the CNOT is the optimum operation to
extract work.

One might ignore the effect of the demon by only considering the reduced
density matrix $\rho _{S}=\mathrm{Tr}_{D}\left[ \rho \right] $ of $S$ by
tracing over the variable of $D$. After the step \emph{S3}, one has the
reduced density matrix $\rho _{S}\left( 3\right) =\mathrm{Tr}_{D}[\rho (3)]$%
. The following thermalization of $\rho _{S}\left( 3\right) $ restores $S$
into its initial equilibrium state $\rho _{S}(1)$ by absorbing heat.
Therefore, the net result of ignoring the demon means that there exists a
perpetual machine of the second kind, which absorbs heat from a single heat
bath and converts it into work. This obvious violation of the second law of
thermodynamics leads to the so-called \textquotedblleft Maxwell's demon
paradox". When the demon is included in the thermodynamic cycle, however,
the \textquotedblleft paradox" disappears and the violation of the second
law is prevented. Hence the present concrete model shows the effect of
Maxwell's demon and verifies the prediction of Landauer's principle.

\emph{Experimental implementation based on superconducting
systems.---} The above Maxwell's demon assisted QHE  model can
be demonstrated by a realistic system, e.g., the superconducting
quantum circuit illustrated in Fig. 2(a), described by the
Hamiltonian~(\ref{0}). Here, two qubits $S$ and $D$ are
specified to two charge qubits~\cite{wang} with the controllable
level spacings $\Delta _{F}=E_{cF}\left\vert
n_{gF}-1/2\right\vert \,\,(F=S,\,D)$,  manipulated by the gate
voltages $V_{gF}$, where $E_{cF}$ is the effective charging
energy and $n_{gF}=V_{gF}C_{gF}/2e$ is the offset reduced gate
charge of the qubit $F$. Here, the magnetic fluxes threading the
two qubits are set to $\Phi _{0}/2$. The coupling constant
$E_{L}=E_{0}\cos (\pi \Phi _{x}/\Phi _{0})$ can be tuned to zero
by the external magnetic flux $\Phi _{x}$ through the dc SQUID
$L$, where $E_{0}$ is the Josephson tunnelling energy.
Therefore, the inter-qubit coupling can be switched on and off
by the magnetic flux $\Phi _{x}$. Below We explain how to
implement our QHE by using the circuit in Fig.~2(a).

\begin{figure}[tbp]
\begin{center}
\includegraphics[bb=70 290 530 680, width=8 cm, clip]{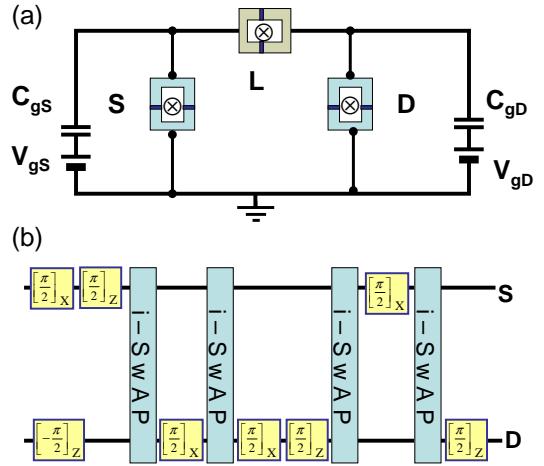}
\end{center}
\caption{(color online). A Maxwell's demon QHE implemented by a
superconducting circuit. (a) two charge qubits $S$ and $D$ with
different localized temperatures function as the working
substance and the demon, respectively. $V_{gF}$ and $C_{gF}$ are
the gate voltage and capacitance of qubit $F$. (b) The quantum
logic operations (two CNOTs) to simulate the demon are realized
by four i-SWAP operations together with several single-bit
operations~\protect\cite{schuch}, e.g., $\left[ \Theta \right]
_{X} $, a $\Theta $-rotation along the $x$ axis. Here $\Theta
=\pm \pi /2$. The quantum control to implement these operations
is carried out by the dc SQUID $L$ in (a).} \label{fig2}
\end{figure}

To implement the step \emph{S1} in our proposal, we turn off the interaction
between two qubits by applying the magnetic flux $\Phi _{x}=\Phi _{0}/2$.
The two charge qubits are coupled to their own baths, which can be realized
by two local temperatures $T_{S}$ and $T_{D}$. This temperature difference
can be guaranteed by temperature gradient. Thus, with a thermalization time
of about $1\sim 10\,\mu $s~\cite{you}, two superconducting charge qubits can
reach their own equilibrium states, described by Eq.~(\ref{1}). The
population of each qubit is described by the steady term of Eq.~(\ref{9}).

After two qubits reach their equilibrium states, the step
\emph{S2} starts to implement a CNOT operation with the demon
being the target qubit, and the working substance being the
controlling qubit. This CNOT can be obtained~\cite{schuch} as
follows: First, two single-qubit operations, $[\pi /2]_{X}$ and
$[\pi /2]_{Z}$, are applied on the system as well as one
single-qubit operation, $[-\pi /2]_{Z}$, on the demon, here
$\left[ \Theta \right] _{i}$ ($i=X,\,Z$) denotes a $\Theta
$-rotation along the $i$ axis. Second, we turn on the two-qubit
interaction by setting $\Phi _{x}=0$ and $n_{gS}\approx
n_{gD}\approx 1/2$. Then, two coupled qubits evolve in time
$t_{0}=\pi \hbar /(4E_{0})$ (about $1\sim 10$ $n$s~\cite{wang})
through the Hamiltonian~(\ref{0}) to get an i-SWAP operation.
Third, turn off the two-qubit interaction, and a $[\pi /2]_{X}$
operation is only applied to the demon. Fourth, the two-qubit
interaction is turn on again and the two qubits evolve
$t_{0}=\pi \hbar /(4E_{0})$. Finally, switch off the two-qubit
interaction, and a $[\pi /2]_{Z}$ operation is applied to the
system. After these steps, a CNOT operation is implemented on
the two qubits by the quantum circuit in Fig.~\ref{fig2}(a) and
Eq.~(\ref{2}) is obtained.

We now consider the step \emph{S3}. In our proposed experimental
setup in Fig.~\ref{fig2}(a), the CEV operation in the step
\emph{S3} is chosen as a CNOT operation, with the demon being
the controlling qubit, and the working substance being the
target qubit. This CNOT can be obtained similarly as the step
\emph{S2}. In this case, the work done by the QHE is maximum,
since the demon flips the system from the excited state
$|1\rangle _{S}$\ to the ground state $|0\rangle _{S}$. After
the step \emph{S3}, the qubit interaction is switched off by
$\Phi _{x}=\Phi _{0}/2$, then our QHE starts a new cycle. The
two CNOT operations used in \emph{S2} and \emph{S3} are
schematically shown in Fig.~\ref{fig2}(b). The total time for
these two operations in the charge-qubit circuits~\cite{wang} is
$\sim 10$ ns, which is much less than the relaxation time $1\sim
10\,\mu $s~\cite{you}.

In the above quantum circuits, if the temperature $T_{D}$ is so
low that $\exp (-\beta _{D}\Delta _{D})\ll 1$, the efficiency
$\eta $ in Eq.~(\ref{6}) of our proposed QHE approaches $\eta
=1-\Delta _{D}/\Delta _{S}$ of a simple quantum Otto heat
engine~\cite{explanation,kieu,quan}. In the experimental setup,
the parameters usually are of the following order of the
magnitude \cite{tsai}: $E_{cS}\sim 10^{-23}$ J, $\left\vert
2n_{gS}-1\right\vert \sim 10^{-2}$, $T_{S}\sim 10^{-2}$ K.
Hence, $\exp (-\beta _{S}\Delta _{S})\sim e^{-1}$. If we choose
$E_{cS}\sim E_{cD}$ and $T_{D}\sim (T_{S}/10)\,\sim 10^{-3}$ K,
then we certainly have $\exp (-\beta _{D}\Delta _{D})\sim
e^{-10}\ll 1$. Using the parameters about $\Delta _{D}$ and
$\Delta _{S}$ of the superconducting qubits, the efficiency
$\eta $ of the QHE\ can be further given by
\begin{equation}
\eta =1-\frac{\left\vert 2n_{gD}-1\right\vert }{\left\vert
2n_{gS}-1\right\vert },  \label{10}
\end{equation}%
which is independent of $T_{S}$ and $T_{D}$. Here, we have
adjusted the macroscopic quantum circuit to be symmetric with
respect to $D$ and $S$ by choosing $E_{cS}=E_{cD}$. For
instance, if $n_{gD}=0.498$ and $ n_{gS}=0.492$, the efficiency
becomes $\eta =0.75$.

The derived expression, in Eq.~(\ref{10}), for the QHE
efficiency could be tested by experiments on superconducting
qubit circuits. There are three important conditions for the
experimental implementation of our QHE: i) controllable
two-qubit operations; ii) two different temperatures $T_{S}$ and
$T_{D}$ for the two nearby qubits; and iii) precise measurements
of the power of the microwave irradiations. The first one has
been discussed above. The second condition could be achieved by
a temperature gradient on the chip. For the third condition, a
precise measurement of the power spectrum of the microwave is
experimentally accessible in these circuits. Hence, the heat $
Q_{\mathrm{in}}$ absorbed by $S$ and the heat $Q_{\mathrm{out}}$
released by $D$ can be measured, when they are in contact with
their respective baths in the step \emph{S1}. Similar to the
arguments in Ref.~\cite{lloyd,scully,kieu,quan}, the work
produced in this cycle depends on the conservation of energy
$W=Q_{\mathrm{in}}-Q_{\mathrm{out}}$ and does not depend on the
specific operation performed. We can also estimate the output
power of the QHE. From Eqs.~(\ref{6}) and ~(\ref{10}), we have
$W\sim \eta Q_{\mathrm{in}}\sim \eta \,\Delta _{S}\,p_{S}(1)\sim
10^{-25}$J, and the time interval of a cycle is about $\tau \sim
10\,\mu $s. Hence the output power becomes $P=W/\tau \sim
10^{-20}$ J s$^{-1}$. We emphasize that we are now interested in
conceptual designs of new types of QHEs,  rather than their
engineering applications.

In summary, we have studied the operation of a
Maxwell's-demon-assisted QHE and justified the predictions of
Landauer's principle: i) a measurement does not necessarily lead
to entropy increase~\cite{Landauer,bennett,zurek}; and ii) the
apparent violation of the second law does not hold when the
restoration of the demon's memory is included in the cycle
~\cite{demon,Landauer,bennett,maruyama,zurek,lloyd,scully},
because under certain conditions, our composite QHE is
equivalent to a simple quantum Otto engine. We also use
superconducting quantum circuits as an example showing how to
implement this QHE.

FN is supported in part by NSA, LPS, ARO, and NSF. CPS is
partially supported by the NSFC and NFRP of China.


\begin{thebibliography}{99}
\bibitem{explanation} For an Otto type QHE~\cite{kieu,quan} using a
potential well (e.g., an infinite square or a harmonic oscillator), the
efficieny of the Otto cycle (two isochoric and two adiabatic processes) is $%
1-\Delta _{2}/\Delta _{1}$, where the level-spacing $\Delta _{1}$ ($\Delta
_{2}$) of the first (second) isochoric process is determinded by the width
of the potential well. Each isochoric process implies that the level spacing
is constant. In our model, the thermalization processes (step $1$) play the
role of the isochoric processes. The efficiency and positive-work condition
of our model becomes that of known engines in~\cite{kieu,quan} under certain
conditions. We also noticed the inconsistency in defining Otto type QHE, see
e.g., Ref. \cite{scully1}, which do not affect the main idea of our paper.
\end{thebibliography}
\end{document}